\documentclass[11pt,twoside]{article}
\usepackage{asp2014}
	\newcommand{\ass}{\alpha_\textrm{ss}}
	\newcommand{\Minj}{\dot{M}_\textrm{inj}}
	\newcommand{\Teff}{T_\text{eff}}
	\newcommand{\Porb}{P_\text{orb}}
	\newcommand{\Vrot}{v_\text{rot}}
	\newcommand{\Vcri}{v_\text{c}}
\aspSuppressVolSlug
\resetcounters

\begin{document}

\title{Radiative transfer on decretion discs of Be binaries}
\author{Despina~Panoglou$^1$, Daniel~M.~Faes$^1$, Alex~C.~Carciofi$^1$, Atsuo~T.~Okazaki$^2$ and Thomas~Rivinius$^3$\\
\affil{$^1$Instituto de Astronomia, Geof\'isica e Ci\^encias Atmosf\'ericas, Universidade de S\~ao Paulo, SP 05508-900, Brazil; \email{panoglou@usp.br}}
\affil{$^2$Faculty of Engineering, Hokkai-Gakunen University, Sapporo, Hokkaido 062-8605, Japan; \email{okazaki@lst.hokkai-s-u.ac.jp}}
\affil{$^3$European Organisation for Astronomical Research in the Southern Hemisphere, Casilla, Santiago 19001, Chile; \email{triviniu@eso.org}}}

\paperauthor{Despina~Panoglou}{panoglou@usp.br}{}{Instituto de Astronomia, Geof\'isica e Ci\^encias Atmosf\'ericas}{Universidade de S\~ao Paulo}{S\~ao Paulo}{SP}{05508-900}{Brazil}
\paperauthor{Daniel~M.~Faes}{moser@usp.br}{}{Instituto de Astronomia, Geof\'isica e Ci\^encias Atmosf\'ericas}{Universidade de S\~ao Paulo}{S\~ao Paulo}{SP}{05508-900}{Brazil}
\paperauthor{Alex~C.~Carciofi}{carciofi@usp.br}{}{Instituto de Astronomia, Geof\'isica e Ci\^encias Atmosf\'ericas}{Universidade de S\~ao Paulo}{S\~ao Paulo}{SP}{05508-900}{Brazil}
\paperauthor{Atsuo~T.~Okazaki}{okazaki@lst.hokkai-s-u.ac.jp}{}{Faculty of Engineering}{Hokkai-Gakunen University}{Sapporo}{Hokkaido}{062-8605}{Japan}
\paperauthor{Thomas~Rivinius}{triviniu@eso.org}{}{European Organisation for Astronomical Research in the Southern Hemisphere}{}{Casilla}{Santiago}{19001}{Chile}

\begin{abstract}
In this work we explore the effect of binarity in the decretion disc of Be stars, in order to explain their variability. To this aim, we performed smoothed particle hydrodynamics (SPH) simulations on Be binary systems, following the matter ejected isotropically from the equator of the Be star towards the base of an isothermal decretion disc.
We let the system evolve for time long enough to be considered at steady state, and focus on the effect of viscosity for coplanar prograde binary orbits. The disc structure is found to be locked to the orbital phase, exhibiting also a dependence on the azimuthal angle.

Additionally, we present the first results from detailed non-local thermodynamic equilibrium (non-LTE) radiative transfer calculations of the disc structure computed with the SPH code. This is achieved by the use of the three-dimensional (3D) Monte Carlo code HDUST, which can produce predictions with respect to a series of observables.
\end{abstract}

\section{Introduction}
Be stars have similar spectra with B stars, but rotate very fast (typically 75\% of the critical rotation speed). They are the most prominent case of main sequence stars surrounded by circumstellar outflowing matter, which is adequately represented by a viscous decretion disc.
Be stars are highly variable, with both short- and long-term variability that can be observed through photometry, spectroscopy and polarimetry.

Many models have attempted to theoretically predict the observed properties of Be stars; all of them are connected to the Be decretion disc. \cite{Owoc06} refers to several mechanisms that can lead to the formation of the disc. For instance, it might be the result of surface explosions or of a stellar wind being compressed to a disc due to the rapid rotation \citep{BjCa93}.
In practice non-radial forces suppress the flow around the equator, which is essential to the formation of the disc \citep{OwCG96}, and therefore this scenario has been quite disproven. Another mechanism is pulsations \citep{RiCa13}; correlation has been reported between photometric pulsation and circumstellar activity, but the amplitude that can be inferred by observations seems insufficient to eject matter.

Varying disc feeding rates set the phases of disc growth and dissipation, and might cause temporal variability.
Periodic variability might be due to either density waves and stellar oscillations, or due to the gravitational interaction with a potential binary companion; the latter usually is the case for less strong variations.
Short-term low-amplitude density waves can also be caused by tidal interaction in (eccentric) binary systems.
The long-term high-amplitude V/R variations exhibited by Be stars can only be explained by global density waves.

In this contribution the variability of Be observables is attributed to the tidal interaction of the decretion disc with a binary companion. Lately we finished a study of the disc structure in coplanar orbits \citep{PCO15}, exploring the parameter space in terms of the Shakura-Sunyaev viscosity parameter for the disc gas \citep{ShSu73}, the orbital period, the binary mass ratio and the eccentricity.
We hereby focus on the effect of viscosity, and discuss how the azimuthal and orbital phase variability of the disc affect the observables, giving insight to the dependence on the observational inclination.

\section{Methodology, tools and simulation parameters}\label{s:meth}
The 3D SPH code \citep{OkBa02} is based on a code introduced by \cite{BaBo95}, and is used to simulate the evolution of the system. The two stars, handled as point masses, as well as the particles ejected isotropically from the equator of the Be star, are followed in their routes and their properties are computed, taking into account all the interactions between them.
In selected epochs it is possible to extract the disc structure to a file that can be imported to the 3D non-LTE Monte Carlo radiative transfer code HDUST \citep{CaBj06}. HDUST is able to calculate the temperature structure of the disc and compute the radiation transfer in order to produce spectroscopic, polarimetric and interferometric observational predictions.

The simulations are for a circular coplanar binary system of an orbital period $\Porb=30$ d. The primary Be star has effective temperature $\Teff=19370$ K, radius $R_1=R_*=5.5R_\odot$ and mass $M_1=11.2M_\odot$. The secondary star has $R_2=10^{-6}R_\odot$ and $M_2=0.94M_\odot$, resulting to a binary mass ratio of \mbox{$q=M_2/M_1=0.08$} and a semi-major axis \mbox{$a=17R_*$}.
The SPH simulations were performed with a constant injection rate of $\Minj=10^{^-7}M_\odot$/yr, starting from a disc-less state. These conditions result to a final number of particles $>20000$ in the end of the simulation with the disc at steady state, and a number density at the disc base of the order $n_0\sim10^{13}\text{cm}^{-3}$.
For the HDUST simulations we assumed a rotational speed $\Vrot=0.8\Vcri$.
The main input parameters and variables at steady state of each simulation are shown in \autoref{t:pars}.

\begin{table}[!ht]
\caption{The input parameters of the SPH simulations presented in this text (viscosity parameter $\ass$; eccentricity $\epsilon$; the number of particles $N_\text{inj}$ into which the total ejected mass per time step is divided), along with the values of some variables that can be estimated a posteriori (time to steady state $t_\text{ss}$; final number of particles $N_\text{sph}$; final number density at the base of the disc $n_0$; total disc mass $M_\text{disc}$).}\label{t:pars}\centering
\smallskip
\begin{tabular}{cccc|ccccccccc}
\hline
\noalign{\smallskip}
$\ass$ & $\epsilon$ & $N_\text{inj}$ & $t_\text{tot}$ & $t_\text{ss}$
                    & $N_\text{sph}$ & $n_0$ & $M_\text{disc}$ \\
       &            &                & ($\Porb$)     & ($\Porb$)
                    &                & ($10^{13}\text{cm}^{-3}$) & ($10^{-9}M_\odot$)     \\
\noalign{\smallskip}
\hline
\noalign{\smallskip}
0.1 & 0.0 &  1000 & 120 & 117 & 52264  & 9.54 & 8.22 \\
0.4 & 0.0 &  5000 &  55 &  42 & 46361  & 7.40 & 3.03 \\
1.0 & 0.0 &  5000 &  30 &  21 & 21822  & 3.87 & 1.43 \\
\noalign{\smallskip}
\hline
\end{tabular}
\end{table}

\section{Disc structure}\label{s:dis}

Because of the variability of Be decretion discs, it is important to constrain the disc state so as to understand the observational features of Be binaries.

A direct effect of binarity comes from the simultaneous exertion of the resonant and the viscous torque on the particles of the surrounding gas. The combination of the two results in the truncation of the Be decretion disc at a distance smaller than the periastron distance \citep{OkNe01a,OkNe01b}, where the density structure is heavily affected by the existence of the binary companion.
Truncation (i)~limits the disc extend that is observationally significant, since the outer disc is dissolved and becomes very sparse, and (ii)~causes the accumulation of matter in the inner region, changing the density fall-off slope in comparison to the disc of an isolated star.

For isolated Be stars it has been established that the disc surface density is a power law of the radial distance $r$ given by the relation $\Sigma(r)=\Sigma_0(r/R_*)^{-m}$, where $\Sigma_0$ is the surface density at the base of the disc, and the exponent is equal to $m=2$ when the system is sufficiently evolved.
This relation cannot be valid for Be discs in binary systems for two basic reasons: (i)~it presumes an azimuthal symmetry, which ignores the dependence of the tidal effect on the position of the secondary, and (ii)~it fails to take into account the truncation of the disc.

The truncation radius practically divides the disc into two radial regions: the inner region where the surface density decreases exponentially with increased distance, with an exponent close to the exponent for isolated stars, and an outer region where the power law exponent is much larger in absolute values, i.e.~the surface density decreases more steeply.
Truncation is taken into account in a relation proposed in \cite{OkBa02} for the \emph{azimuthally averaged} surface density of the decretion disc in Be binaries:
  \begin{equation}
  \left<\Sigma(r)\right>_\phi
  = A\frac{(r/R_t)^{-m}}{1+(r/R_t)^{n-m}}\label{e:Atsuo}
  \end{equation}
where $R_t$ is the azimuthally averaged truncation radius, i.e.~the distance that separates the inner and outer parts of the disc; $m$ and $n$ are the exponents of the inner and outer parts of the disc, respectively; and $A$ is a constant connected to the surface density at the base of the disc.

\cite{PCO15} gave a more general form of this relation that takes into account both the azimuthal and the phase dependence of the disc structure:
  \begin{equation}
  \Sigma(r,\phi,p)
  = A(\phi,p)\frac{(r/R_{t}(\phi,p))^{-m(\phi,p)}}
  {1+(r/R_{t}(\phi,p))^{n(\phi,p)-m(\phi,p)}}\label{e:Atsne}
  \end{equation}
where the four parameters of eq.~\eqref{e:Atsuo} are now themselves functions of the azimuthal angle $\phi$ (\deg) and the orbital phase $p\in[0,1)$ ($p=0$ at apastron, $p=0.5$ at periastron, when in eccentric orbits).
After an adequately long time for the formation and evolution of a Be binary, a disc can be formed around the Be star and reach a ``quasi-steady state'', in the sense that there are no substantial variations between the same orbital phase of subsequent cycles, so that eq.~\eqref{e:Atsne} is valid no matter the orbital cycle.

\cite{PCO15} showed that in coplanar circular systems the disc structure is generally in phase with the secondary's orbit. The azimuthal dependence of the disc parameters ($R_t,m,n,A$) for different phases is only shifted in the azimuthal direction by an amount proportional to the phase difference. Mathematically this can be expressed as $x(\phi,p+\Delta p)=x(\phi+\Delta\phi,p)$, with $\Delta\phi=\Delta p\cdot360\deg$ and $x\in\{R_t,m,n,A\}$.
This means that, after the steady state has been reached, the decretion disc in circular binaries does not change in shape, but it rotates in phase with the secondary, as shown in \autoref{f:csd}.

\articlefigure[scale=.5]{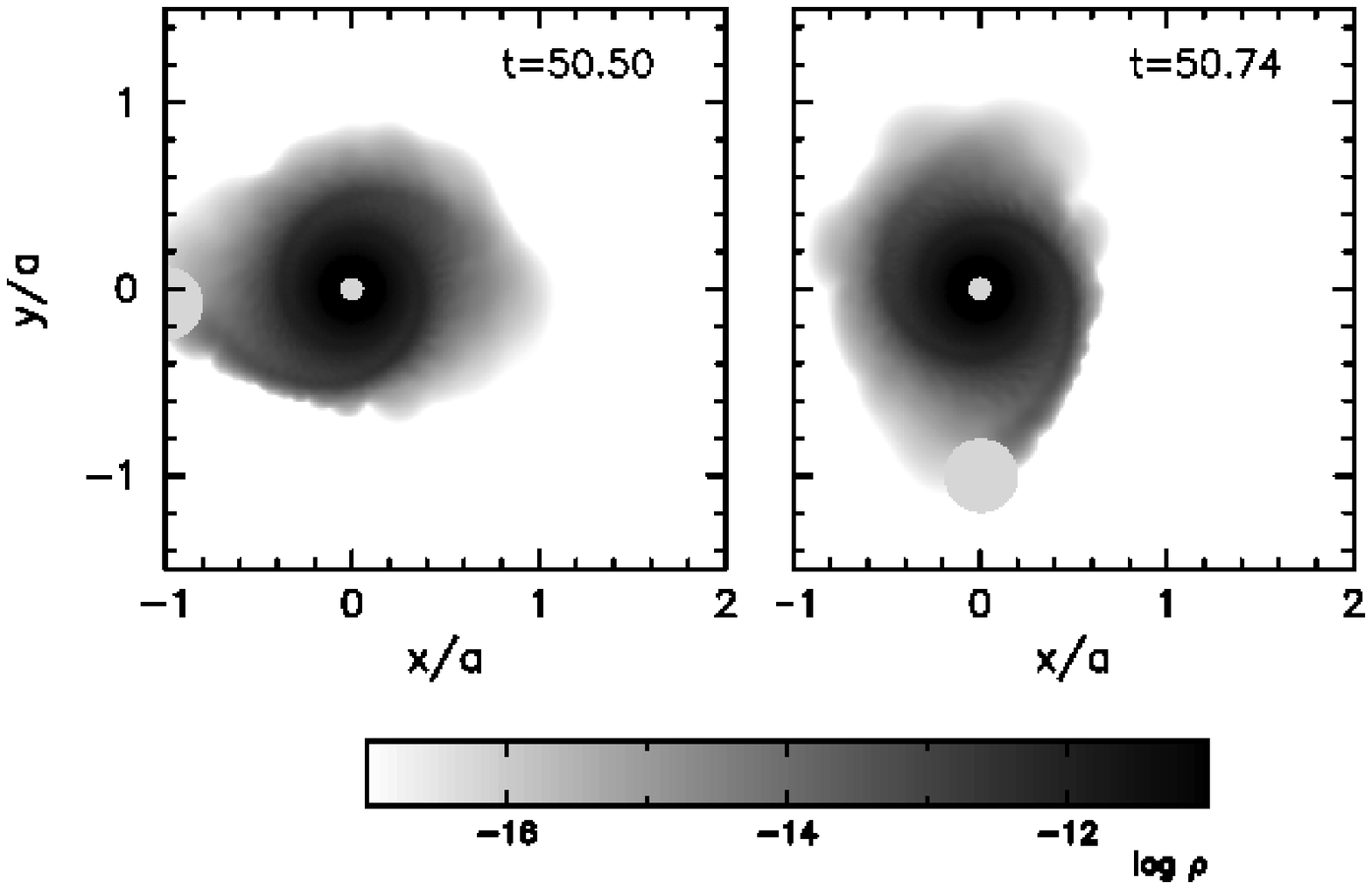}{f:csd}{Surface plots of the decretion disc density for a circular binary system with $\ass=0.4$ at two different orbital phases, after it has reached a quasi-steady state. The stars are designated as gray filled circles, with the length scale centered at the Be star. The time $t$ from the beginning of the simulation is given in units of $\Porb$ for each snapshot. Adopted from \cite{PCO15}.}

\section{Observational predictions}
As mentioned in \autoref{s:meth}, it is possible to extract the disc structure calculated with the SPH code at some epoch into a file that can be imported to the radiative transfer code HDUST. In turn, HDUST computes the temperature structure of the designated disc configuration, and then calculates the radiation flux at various wavelengths in the desired positions of observers.

An example of the result of this procedure is shown in \autoref{f:flux} for different values of the disc viscosity. The flux is higher for lower viscosities. This is a consequence of the fact that low viscosity discs are more dense (\autoref{t:pars}), which results to higher flux for wavelengths $\lambda>0.365\mu$m (Balmer jump):
With the inner disc having higher density, more light is absorbed in $\lambda<0.365\mu$m and this energy is re-emitted in longer wavelegths, producing this flux excess.

\articlefigure{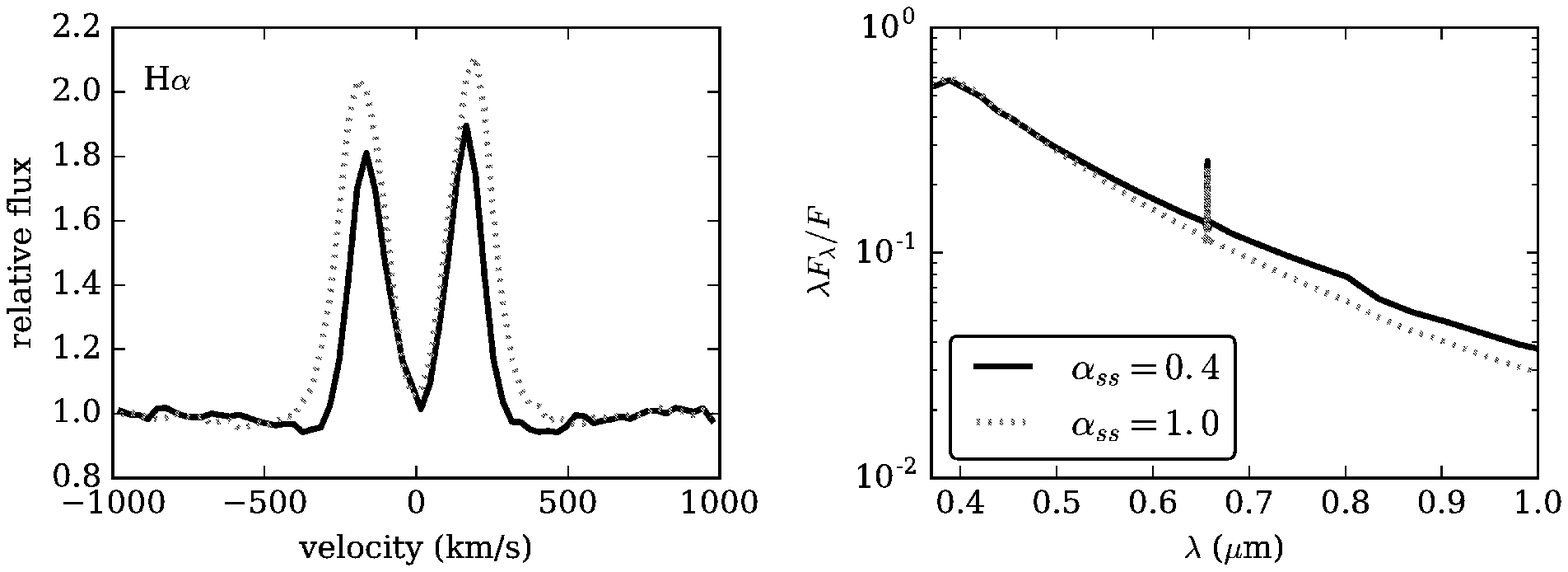}{f:flux}{Emergent spectrum of the disc in two binary systems of different viscosity for an observer at position $(\phi,i)=(0,75)\deg$ at phase $p=0$, after the systems have reached the steady state. Left: H$\alpha$ line profile; right: Visible spectral energy distribution.}

\articlefiguretwo{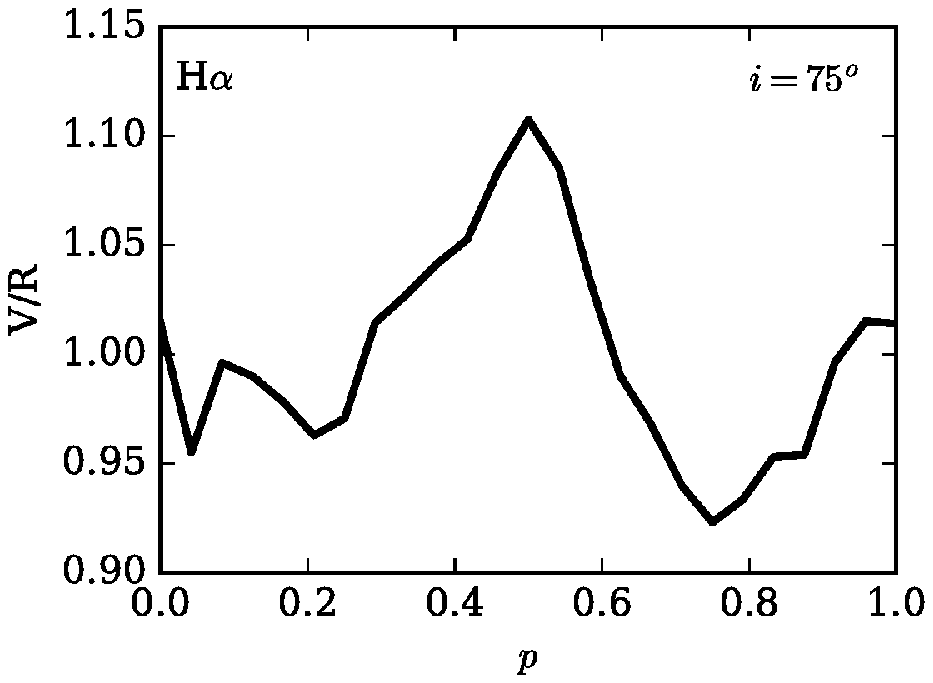}{ats1-2a-0a}{f:VoR}{Circular Be binary system. Left: The H$\alpha$ V/R ratio as a function of the orbital phase for a system with disc viscosity $\ass=0.4$, seen with inclination $75\deg$. Right: The truncation radius as a function of the azimuthal angle for $\ass=0.1,1.0$.
The filled circle shows the position of the secondary, and the error bars correspond to the validity of the fittings of eq.~\eqref{e:Atsne} to the surface density of each azimuthal angle $\phi$. In order to minimize the errors, the surface density of each $\phi$ was folded for 5 cycles at steady state.}

In order to compute the time variability of any observable for a binary system, HDUST should run multiple times within the orbital cycle. But as explained in \autoref{s:dis}, the variability in time for circular systems can be translated to variability in terms of the azimuthal angle $\phi$. Hence it was possible to calculate the V/R ratio as a function of the orbital phase $p$ (\autoref{f:VoR}, left panel).
The right panel of \autoref{f:VoR} confirms a similar variability for the truncation radius, as computed by fitting the parameters of eq.~\eqref{e:Atsne} for a number of equally spaced azimuthal angles.

Note that the maximal truncation radius occurs with some phase difference with respect to the direction of the secondary, varying from about $10\deg$ for $\ass=1$ to about $60\deg$ for $\ass=0.1$. The maximum of V/R occurs with a lag of about $\pi$.

\section{Discussion and perspectives}
We showed the first results of observational predictions prepared by calculating the radiation transfer in Be disc structures computed separately with an SPH code. We found that for an inclination angle $i=75\deg$ the emerging V/R ratio exhibits a variability in the same time scale as the orbital one.

It is suggested that the appearance of variability for various observables, apart from varying with the wavelength (range) under study, is strongly connected to the inclination angle of the observer. Seeing the star edge-on should give a more intense V/R variability. The two high peaks of the V/R ratio as a function of the orbital phase are expected to be almost equal for low viscosities, resulting in variability cycles of periods equal to $\Porb/2$. On the contrary, seeing the star pole-on should not allow for countable variations.
Those hypotheses and others remain to be verified.

This first phase of our study already allowed us to compute the V/R variation along the orbital cycle, but a complete exploration of various systems and observers' angles is necessary to get a full understanding of how binarity affects the properties of Be stars. The symmetry break in the line profiles, measured as V/R variations, and/or other observables may be used as a tracer of the system properties.

\acknowledgements
DP thanks D.~Baade for the fruitful discussions that helped improve understanding of observational expectations during the days of the conference.
This work made use of the computing facilities of the Laboratory of Astroinformatics (IAG/USP, NAT/Unicsul), whose purchase was made possible by the Brazilian agency FAPESP (grant \mbox{2009/54006-4}) and the \mbox{INCT-A}. DP acknowledges support from FAPESP (grant \mbox{2013/16801-2}). DMF acknowledges support from CNPq (grant 200829/2015-7) and FAPESP (grant 2016/16844-1).
ACC acknowledges support from CNPq (grant \mbox{307594/2015-7}) and FAPESP (grant \mbox{2015/17967-7}).

\bibliography{biblio}

\end{document}